\documentclass[twocolumn, amsmath,superscriptaddress, amsfonts,prb]{revtex4}
\usepackage{graphicx}
\begin{document}

\title{Electrostatic theory of metal whiskers}
\author{V. G. Karpov}\email{victor.karpov@utoledo.edu}\affiliation{Department of Physics and Astronomy, University of Toledo, Toledo, OH 43606, USA}
\begin{abstract}
Metal whiskers often grow across leads of electric equipment and electronic package causing current leakage or short circuits and raising significant reliability issues. The nature of metal whiskers remains a mystery after several decades of research. Here, the existence of metal whiskers is attributed to the  energy gain due to electrostatic polarization of metal filaments in the electric field. The field is induced by surface imperfections: contaminations, oxide states, grain boundaries, etc. A proposed theory provides closed form expressions and quantitative estimates for the whisker nucleation and growth rates, explains the range of whisker parameters and effects of external biasing, and predicts statistical distribution of their lengths.

\end{abstract}

\date{\today}

\maketitle
\section{Introduction}\label{sec:intro}
\begin{figure}[b!]
\includegraphics[width=0.5\textwidth]{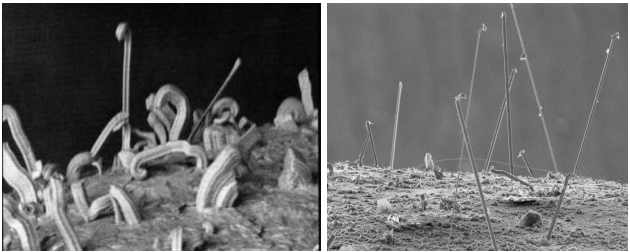}
\caption{SEM pictures of tin (left) and zinc (right) whiskers. Reproduced from the NASA photogallery. \cite{photo} \label{Fig:whiskimage}}
\end{figure}

Metal whiskers are hair-like protrusions observed at surfaces of some metals; tin and zinc examples are illustrated in Fig. \ref{Fig:whiskimage}. \cite{nasa1,calce,nasa2,bibl,nasa3,photo} In spite of being omnipresent and leading to multiple failure modes in electronic industry, the mechanism behind metal whiskers remains unknown after more than 60 years of research. While, some consensus, at a qualitative level, is that whiskers represent a stress relief phenomenon, that never led to any quantitative description including order-of-magnitude estimates of whisker parameters. A  theory of metal whiskers presented here is consistent with many published observations and provides quantitative analytical results.
\bigskip

As a brief survey of relevant data, \cite{nasa1,calce,nasa2,bibl,nasa3,photo,shetty2004,crandall,munson,fang2006} it should be noted that whiskers grow up to $\sim 1-10$ mm in length and vary from $\sim 1$ nm to $\sim 30$ $\mu$m in diameter. Their parameters are characterized by broad statistical distributions: side by side with fast growing whiskers there can be others, on the same surface, whose growth is much slower or completely stalled. The metal surface conditions play a significant role. In particular, oxide structure and various contaminations are important factors determining whisker concentration, growth rate and dimensions. The metal grain size appears to be less significant for small grains (nanometers to few microns), while whiskers are unlikely for very large grains; recrystallization can be of importance. \cite{sarabol2013}  Various additives can have significant effects on whisker growth, such as e. g. small concentration of Pb strongly suppressing tin whiskering. Electric bias  was reported to exponentially increase whisker growth rate, \cite{crandall,liu2004,lin2008} which was attributed to the effects of electric current (a possible role of the electric field was not analyzed), although the opposite statements have been made recently. \cite{ashworth2013} A common observation is that whiskers grow from the root rather then from the tip, and the material required for their growth is supplied from large distances through long range surface diffusion rather than from a narrow neighboring proximity; there is no surrounding dent formed in the course of whisker growth.  A comprehensive review of experimental data on the most studied tin whiskers before the year of 2003 was given in a monograph \onlinecite{galyon2003}; Refs. \onlinecite{choi2004, zhang2004, tu2005, bunian2013} provide more recent summary.

Multiple attempts to understand the mechanisms of whiskers growth (see e.g. Refs. \onlinecite{nakai2009,sobiech2008,smetana2007,barsoum2004}) revolved around the role of surface stresses relived by whisker production, dislocation effects, and oxygen reactions. It was shown \cite{lindborg1976,tu2005} that stress gradients along with certain assumption about system parameters can explain tin whisker growth rates but not their existence, shapes and statistics. Overall, these attempts have not lead to verifiable quantitative predictions.

The mechanism proposed here is qualitatively different as driven by the existence of strong enough electric field $E$ above the metal surface. The field is generally due to surface imperfections, such as oxide, ion contaminations, local stresses, and interfacial states. The appearance of whiskers is described as the electric field induced nucleation. It is triggered by the energy gain $F_E=-{\bf p\cdot E}$ due to the induced whisker dipole ${\bf p}=\alpha {\bf E}$ in the electric field $E$,  where $\alpha$ is the polarizability. The latter is anomalously strong for the needle shaped metallic particles that serve as whiskers' nuclei.

This paper is organized as follows. Sec. \ref{sec:fin} describes the field induced nucleation of whiskers estimating their initial  dimensions and nucleation rates. Random electric fields responsible for whisker nucleation are described in Sec. \ref{sec:field}; they turn out to be different at small, intermediate, and large distances. The kinetic of whisker growth is analyzed in Sec. \ref{sec:growth}. Sections \ref{sec:longr} and \ref{sec:stat} discuss respectively the phenomenon of long-range surface diffusion related to whisker growth and statistical distributions of whisker lengths. Numerical estimates in Sec. \ref{sec:num} prove that the proposed theory is in at least semi-quantitative agreement with the data. The conclusions in Sec. \ref{sec:concl} list this theory successes and  problems.

\section{Field induced nucleation}\label{sec:fin}
\begin{figure}[t!]
\includegraphics[width=0.4\textwidth]{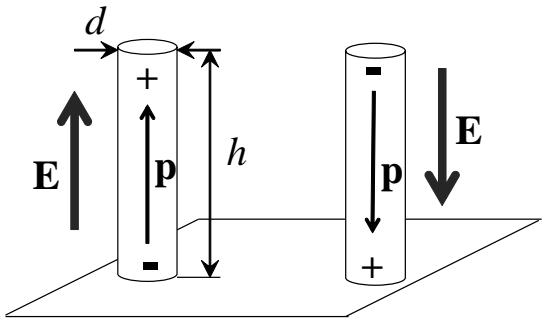}
\caption{Sketch of two whiskers of length $h$ and diameter $d$ on a metal surface with local electric fields ${\bf E}$ (of opposite directions) inducing the dipole moments ${\bf p}$. \label{Fig:whiskgeom}}
\end{figure}

More exactly, the electrostatic energy gain in the electric field can be represented as \cite{kaschiev2000,warshavsky1999,isard1977}
\begin{equation}F_E=-\varepsilon\alpha  E^2 \label{eq:Egain}\end{equation}
where $\varepsilon$ is the dielectric permittivity of the surrounding medium. This energy gain is independent of the sign of the electric field, outwards or towards the surface, as illustrated in Fig. \ref{Fig:whiskgeom}. In this section, we consider field induced nucleation in a uniform field. Complications due to the field nonuniformity will be discussed in Sec. \ref{sec:field}.
\bigskip

The polarizability $\alpha$ is a maximum in the longitudinal direction illustrated in Fig. \ref{Fig:whiskgeom}; it is by approximately the factor of $(h/d)^2\gg 1$ greater than the particle volume $\pi (d/2)^2h$ that serves as a standard measure of polarizability in electrostatics. The mechanism of that enhancement can be understood as follows. Under electric field $E$, a metallic needle will accumulate at its ends opposite charges of absolute values $q\sim Eh^2$, just sufficient to cancel out the field inside the needle. They correspond to the dipole moment $p\sim qH \sim Eh^3 \sim EV(h/d)^2$ where $V\sim hd^2$ is the particle volume.

A more accurate result for needle polarizability (Ref. \onlinecite{landau1984}, p. 17) is given by
\begin{equation}\alpha\approx \frac{h^3}{3\Lambda} \quad {\rm with} \quad \Lambda\equiv \ln(4h/d)-7/3.\label{eq:ncyl}\end{equation}
Strictly speaking, Eq. (\ref{eq:ncyl}) is valid for the case of extremely anisotropic particles with $\Lambda \gg 1$; in what follows, it serves as a simple approximation for even moderately anisotropic structures with $\Lambda\gtrsim 1$.  Note that the concept of energy gain in Eqs. (\ref{eq:Egain}) and (\ref{eq:ncyl}) has been successfully used to describe the field induced nucleation of metal particles.\cite{karpov2007,karpov2008,karpov2008a,karpov2008b,nardone2009,nardone2012,nardone2012a,karpov2012,karpov2012a}

Side by side with the electrostatic energy gain, there is energy loss due to increase in the surface area, $F_S=\pi dh\sigma$  where $\sigma$ is the surface tension. The total change in free energy due to whisker formation is given by
\begin{equation}\label{eq:F}
F(h)=-\frac{h^3}{3\Lambda} \varepsilon E^2 +\pi dh\sigma.
\end{equation}
It is a maximum,
\begin{equation}\label{eq:nucbar}
\max{F(h)}= W(E)\equiv \frac{2}{3}\pi\sigma d\sqrt{\frac{\pi\sigma\Lambda d}{\varepsilon E^2}}
\end{equation}
when
\begin{equation}
h=h_0(E)\equiv\sqrt{\frac{\pi\sigma\Lambda d}{\varepsilon E^2}}.
\label{eq:nuclen}\end{equation}
Here we have treated a logarithmically weak dependence $\Lambda (h)$ as a constant.

\begin{figure}[t!]
\includegraphics[width=0.35\textwidth]{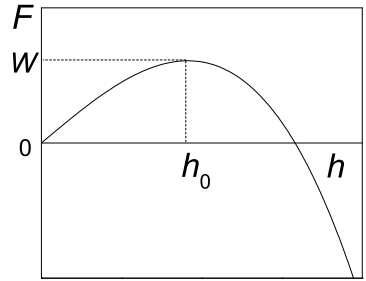}
\caption{Free energy of a whisker vs. its length. \label{Fig:F}}
\end{figure}

The barrier $W$ and its corresponding length $h_0$ shown in Fig. \ref{Fig:F} have the same meaning as the nucleation barrier and radius in the classical nucleation theory. \cite{kaschiev2000} In particular, a whisker becomes stable and keeps growing when its length exceeds $h_0$, so it overcomes the barrier.

Along the lines of standard nucleation theory, the characteristic nucleation time is given by
\begin{equation}\label{eq:nuctime}
\tau = \tau _0\exp\left(\frac{W}{kT}\right).
\end{equation}
Here, the preexponential $\tau _0$ remains rather poorly determined in the framework of the existing classical nucleation theory, possibly leading to many order of magnitude deviations from the data; \cite{james1985,zanotto1985,weinberg1992,kelton1991} its often used values are ranging in the interval $\tau _0\sim 10^{-13}-10^{-8}$ s.

As seen from Eq. (\ref{eq:nucbar}), the nucleation barrier $W$ is field dependent. Since the field is a random variable, the nucleation times are distributed in the exponentially broad interval. One other immediate prediction is that external fields (added to the existing random fields) can exponentially accelerate whisker nucleation.

In Eqs. (\ref{eq:F})-(\ref{eq:nuclen}), we have been tacitly assuming a certain diameter $d$ of the needle shaped nucleus. Both the nucleation barrier and length decrease as $d$ decreases; hence, the smallest possible $d$ are the most favorable. Realistically, $d$ must be greater than some minimum value $d_0$ determined by extraneous requirements, such as sufficient conductivity to support a large dipole energy or sufficient mechanical integrity. Based on data for other types of systems undergoing field induced nucleation, it was estimated that a reasonable minimum diameter is in the sub-nanometer range. \cite{karpov2007,karpov2008,karpov2008a,karpov2008b,nardone2009,nardone2012,nardone2012a,karpov2012,karpov2012a} Lacking more concrete information, we assume here $d_0\lesssim 1$ nm. In the region $d>d_{min}$, the free energy is substantially larger than described by Eq. (\ref{eq:F}), since the energy reducing effect of the electric field cannot be manifested by such thin particles. The region $d=d_0$  can be approximated by a potential wall. Following substantiation in the previous work,\cite{karpov2007,karpov2008,karpov2008a,karpov2008b,nardone2009,nardone2012,nardone2012a,karpov2012,karpov2012a}  we consider nucleation along
the path $d=d_0$; alternative paths that start from the origin introduce only insignificant numerical factors. Correspondingly, we will use $d\lesssim 1$ nm in the above Eqs. (\ref{eq:F})-(\ref{eq:nuclen}) as a rough approximation.

Using the latter, Eqs. (\ref{eq:nucbar}) and (\ref{eq:nuclen}) enable one to estimate the nucleation barrier and length. The energy $\sigma$ in these equations depends on which type of surface is essential. Considering for example tin whiskers, the macroscopically averaged value \cite{alchagirov2007, rice1949} is $\sigma \sim 500$ dyn/cm, while the grain boundary related values can be as low as, \cite{aust1951} $\sigma \sim 100$ dyn/cm, and even \cite{saka1988} $\sigma =30$ dyn/cm. Along with approximations $\Lambda \sim \varepsilon \sim 1$, and $E=1$ MV/cm, these values yield $W\sim 1-10$ eV and $h_0\sim 10$ nm. Such $W$ are in the ballpark of nucleation barriers known for various processes. \cite{kaschiev2000} Also, the underlying assumption of $h/d\gg 1$ is justified by the latter result for $h_0$. Therefore, the field induced nucleation appears a conceivable mechanism of metal whisker conception.

\begin{figure}[t!]
\includegraphics[width=0.40\textwidth]{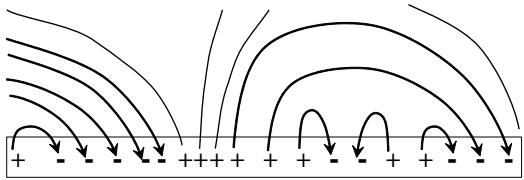}
\caption{Sketch of the the electric field lines in a system of randomly charged patches on a metal surface.\label{Fig:EE}}
\end{figure}

In the existing literature, \cite{cheng2011,nakai2009} the terminology of whisker nucleation  was used as a qualitative statement discriminating between the stages of whisker conception and subsequent evolution. The approach in Eqs. (\ref{eq:F}) - (\ref{eq:nuclen}) provides a quantitative basis for the concept of nucleation. It is triggered by the energy gain of metal whisker due to their polarization  in the surface electric field.

\section{The electric field distribution}\label{sec:field}
Sufficient electric fields above metal surface can arise from spatial variations of the work function. \cite{camp1991} The regions of different surface potential (patches) may be due to the polycrystallinity of the metal: the work function will vary between regions of specific grain orientations by typically a few tenths of a volt. Patch structure may also arise from the presence of adsorbed elements and compounds. Certain features of surface morphology, particularly, its roughness, may result in the electron redistribution. They can be caused by dislocations, \cite{yin2007} stress induced spots of different structure phases, \cite{namahoot2004} or general electric deformation coupling \cite{kolodii2000} in combination with stress induced buckling. \cite{balakrisnan2003,wertheim1991} Local charges due to stress induced oxide cracking or ion trapping under the whisker growing  layer (say, Sn on Cu substrate) are conceivable sources of the above considered surface electric fields as well. Therefore, a surface, that is ideally electrically uniform, may acquire electric surface structure, as illustrated in Fig. \ref{Fig:EE}.

\begin{figure}[t!]
\includegraphics[width=0.35\textwidth]{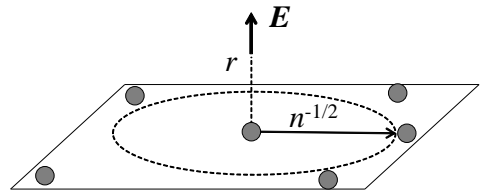}
\caption{The perpendicular electric field component ${\bf E}$  at small distance $r$ from the metal surface. Gray circles represent point surface charges with the average inter-charge distance $n^{-1/2}>r$\label{Fig:EEE}}
\end{figure}

The measurements  reveal indeed the work function fluctuations of $\sim 0.5$ eV induced by $L\sim 10$ $\mu$m patches in some metals. \cite{camp1991} In general, the charged surface state concentration of $n\gtrsim 10^{12}$ cm$^{-2}$ not unusual for many materials \cite{sze} would correspond to the field strength of
\begin{equation}\label{eq:E0}E=E_0\equiv 4\pi en /\varepsilon\gtrsim 10^6\quad {\rm V/cm} \end{equation}
where $e$ is the electron charge. This field can be directed either up or downwards in Figs. \ref{Fig:EE} and \ref{Fig:E} extending over the characteristic distance $L$ above the surface. The above simple expression fails in the regions of both short and large distances as explained next.

\subsection{High charge densities, short nuclei}\label{sec:highch}
The applicability of Eq.(\ref{eq:E0}) is limited by the condition of continuous charge distribution $n\pi r^2\gg 1$ where $r$ is distance above the surface. As applied to the nucleation theory in Sec. \ref{sec:fin}, this implies large enough nucleus height, $nh_0^2\gg 1$, since $h_0$ was derived under the assumption of uniform field $E=E_0$ created by a continuous surface charge distribution. With Eq. (\ref{eq:nuclen}) in mind, this leads to the condition
\begin{equation}\label{eq:nc}
n<n_c\equiv\frac{\sigma\Lambda\varepsilon d}{16 e^2}.\end{equation}
If the surface charge density $n$ exceeds $n_c$, then the embryo height predicted by Eq. (\ref{eq:nuclen}) turns out to be shorter than the inter-charge distance $n^{-1/2}$, and the approximation of uniform field fails.

In the opposite limiting case, $n\gg n_c$, the perpendicular field component (Fig. \ref{Fig:EEE}) can be approximated by the sum of contributions from the nearest {\it point charge} $e$ and all other charges described in continuous approximation by the density $ne$,
\begin{equation}\label{eq:Er}
E\approx\frac{e}{\varepsilon r^2}+\int _{n^{-1/2}}^{\infty}\frac{2\pi nerxdx}{(x^2+r^2)^{3/2}}=\frac{e}{\varepsilon r^2}+E_0r\sqrt{n}\end{equation}
where $E_0$ is given by Eq. (\ref{eq:E0}). For the purely Coulomb contribution, the field induced polarization charge in a metal embryo would be confined to short distances; hence, the embryo height does not contribute to the electrostatic energy gain. (This can be more rigorously proven by modification of the known treatment of the uniform field polarization in Ref. \onlinecite{landau1984} described in the Appendix below.)

The second term in Eq. (\ref{eq:Er}) creates the energy gain similar to that in Eq. (\ref{eq:F}). The difference is that now the acting field is linear in height $h$ due to its proportionality to $r$ in Eq. (\ref{eq:Er}). Therefore, the electric field $E$ in Eq. (\ref{eq:F}) should be replaced with $E_0hn^{-1/2}/2$ (averaged over the embryo length), which makes the electrostatic energy gain proportional to $h^5$. As a result, the nucleation barrier for short embryos in systems with $n>n_c$ can be presented in the form
\begin{equation}\label{eq:shortbar} W=W(E_0)\left(\frac{n}{n_c}\right)^{1/4}, \quad {\rm when}\quad n>n_c\end{equation}
where $W(E_0)$ is defined in Eq. (\ref{eq:nucbar}) with $E=E_0$.

Note that Eq. (\ref{eq:shortbar}) predicts $W\propto n^{-3/4}$, since $W(E_0)\propto E_0\propto n$. Eq. (\ref{eq:nucbar}) describes how the nucleation barrier decreases with surface charge density slightly slower than predicted by the baseline model of the uniform field in Eq. (\ref{eq:nucbar}). Because $(n/n_c)^{1/4}$ is a relatively weak dependence derived in otherwise rather approximate theory, for the sake of simplicity, we will set the latter factor to unity, thus using, in what follows, Eq. (\ref{eq:nucbar}) in the entire range of surface charge densities.

\subsection{Large distances, random fluctuations}\label{sec:largedis}
\begin{figure}[b!]
\includegraphics[width=0.33\textwidth]{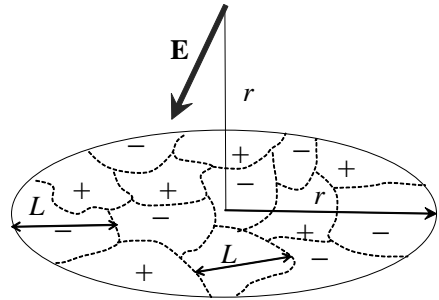}
\caption{Sketch of the patched area domain of radius $r$ where $+$ and $-$ represent positively and negatively charged patches (shown in dash) of characteristic linear dimension $L$ each. The fat arrow represents the random field vector at distance $r$ along the domain axis.\label{Fig:E}}
\end{figure}

As illustrated in Fig. \ref{Fig:E}, at distances $r\gg L$, the contributions of oppositely charged patches mostly cancel each other, and the field is due to a random excess number $\Delta N$ of the patches of a certain sign close enough to the point of observation. Taking the latter at height $r$ above the surface, the charged patches in a domain of radius $r$ beneath will generate more or less perpendicular random field. Therefore, $\Delta N\sim \sqrt{N}\sim r/L$, where $N\sim r^2/L^2$ is the average number of patches in the domain of radius $r$. Hence, one can estimate the absolute value of the field as
\begin{equation}\label{eq:Eest}
E\sim \frac{\Delta Nne L^2}{r^2}\sim E_0\frac{L}{r}.
\end{equation}
The corresponding contribution to free energy is then given by [cf. Eqs. (\ref{eq:Egain}) and (\ref{eq:ncyl})]
\begin{equation}\label{eq:WE1}
F_E=- \frac{\varepsilon E_0^2hL^2}{3\Lambda}\quad {\rm when}\quad h\gg L .
\end{equation}

The angular distribution of field ${\bf E}$ at these distances becomes quite random. The maximum angle of 90 $^{\circ}$ between ${\bf E}$ and the surface normal corresponds to the case when the  positive and negative random charges are as far away from each other as possible (i. e. on the opposite halves of the circle in Fig. \ref{Fig:E}); hence, the angular distribution ranging from 0 to 90 $^{\circ}$ with a maximum somewhere in between.

The field will vary not only in the lateral directions, but with the distance to the surface above any given spot as well. This happens because the contributing random charge configurations change with that distance. Based on the interpretation in Fig. \ref{Fig:E}, the field $E$ will change considerably over distances of the order of $r$. In particular, the characteristic length of fluctuations increases linearly with $r$. It should be understood however that the angular distribution of fields can be significantly different from the observed angular distribution of whiskers. The latter is determined by the kinetic of growth of crystalline structures in random fields, which non-trivial problem is beyond the present scope.

Far enough from the surface, the field in Eq. (\ref{eq:Eest}) becomes very low giving up to the background (thermal) electric field. Its time average is given by
\begin{equation}
\langle E_T^2\rangle =4\pi\sigma _{SB}T^4/c\sim 20\quad {\rm V}^2{\rm cm}^{-2}.
\label{eq:Esq}\end{equation}
Here $\sigma _{SB}$ is the Stefan-Boltzmann constant, $c$ is the speed of light, and we chose the temperature $T\sim 300$ K. Comparing the results in Eqs. (\ref{eq:Eest}) and (\ref{eq:Esq}) yields the overplay distance
\begin{equation}\label{eq:rc}
r_c\sim L\frac{E_0}{E_T}.
\end{equation}
For the above used numerical values, $r_c\sim 10$ cm is far beyond the whiskers length domain. However it falls into that domain, shrinking to e. g. $r_c\sim 0.1$ mm, for the case of high enough temperatures, low surface charge densities, or small patches, say, $L\sim 100$ nm. Since thermal radiation is polarized parallel to the surface, \cite{rytov} it is expected that whiskers in that region will evolve mostly parallel to the surface as well.

\section{Whisker growth}\label{sec:growth}

\subsection{General formalism}\label{sec:genfor}
Whisker growth occurs through the process of many elemental acts of accretion. Such multi-step processes are described by the Fokker-Planck approach (see e.g. pp. 89, 90, and 428 in Ref. \onlinecite{landau2008}) with statistical distribution $f(h,d)$, such that $f(h,d)dhdd$ is the number of whiskers with height and diameter in the intervals $(h,h+dh)$ and $(d,d+dd)$ respectively.  The Fokker-Planck equation takes the form
\begin{equation}\label{eq:FP}
\frac{\partial f}{\partial t}=-\frac{\partial s_h}{\partial h} - \frac{\partial s_d}{\partial d}
\end{equation}
Here, $s_h$ and $s_d$ are the components of the flux in the whisker dimensions space (s$^{-1}$ cm$^{-3}$),
\begin{eqnarray}\label{eq:flux}
s_h&=&-A_hf-B_{hd}\frac{\partial f}{\partial d}-B_{hh}\frac{\partial f}{\partial h}\label{eq:a}\\
s_d&=&-A_df-B_{dh}\frac{\partial f}{\partial h}-B_{dd}\frac{\partial f}{\partial d}\label{eq:b}
\end{eqnarray}
The kinetic coefficients are defined as follows,
\begin{eqnarray}\label{eq:difcoe}
A_h=\tilde{A}_h+\frac{\partial B_{hd}}{\partial d}, &\quad &A_d=\tilde{A}_d+\frac{\partial B_{dh}}{\partial h},\nonumber\\
\tilde{A}_h=\sum _i\delta h_i/t,&\quad &\tilde{A}_d=\sum _i\delta d_i/t,\nonumber \\
B_{hh}=\sum _{ij}\delta h_i\delta h_j/\delta t, &\quad &B_{dd}=\sum _{ij}\delta d_i\delta d_j/\delta t,\\
B_{hd}=B_{dh}&=&\sum _{ij}\delta h_i\delta dj/\delta t\nonumber
\end{eqnarray}
where $\delta h_i$ and $\delta d_i$ are random changes in $h$ and $d$ at a step $i$ in the course of whisker growth over time $\delta t$.

Two boundary conditions to Eq. (\ref{eq:FP}) are $f(r=0)=0$ and $f(r=\infty )=0$. They reflect the facts that very thin filaments cannot exist due to certain extraneous limitations such as loss of conductivity or mechanical instability, and that only finite radii are achievable over finite times $t$.

The approximation of independent height and diameter evolution below, means $B_{hd}=B_{dh}=0$. To further simplify the analysis, the two remaining coefficient are set equal, $B_{hh}=B_{dd}\equiv B$. Relaxing these limitations leads to more cumbersome results without new qualitative features.

The $A$ coefficients are connected with $B$ by a relationship which follows from the fact that $s_h=s_d=0$ for the equilibrium distribution $f_0(h,d)\propto \exp[-F(h,d)/kT]$, where $F$ is the free energy. This yields
\begin{equation}\label{eq:ss}
s_{h}=-Bf_0\frac{\partial }{\partial h}\left(\frac{f}{f_0}\right),\quad s_{d}=-Bf_0\frac{\partial }{\partial d}\left(\frac{f}{f_0}\right)
\end{equation}

Using Eq. (\ref{eq:ss}) for $s$, multiplying Eq. (\ref{eq:FP}) by $h$, integrating from 0 to $\infty$ by parts, and noting that $\int fhdh=\langle h\rangle$, yields $\partial \langle h\rangle /\partial t=\langle\partial F/\partial h\rangle$ (angle brackets denote averages). Neglecting fluctuations in the ensemble of nominally identical filaments enables one to approximate $\langle F\rangle = F(\langle h\rangle )$ and $\langle\partial F/\partial h\rangle=\partial \langle F\rangle /\partial \langle h\rangle$. Similar transformations apply to the $d$ dependence. Omitting for brevity the angular brackets, one finally obtains,
\begin{equation}\label{eq:mobility}
\frac{\partial h}{\partial t}=-b\frac{\partial  F}{\partial h},\quad \frac{\partial d}{\partial t}=-b\frac{\partial  F}{\partial d}\quad {\rm with}\quad b=\frac{B}{kT}.
\end{equation}
These equations have the standard meaning of the relations between the (growth) velocities and the (thermodynamic) forces $-\partial F/\partial h$, $-\partial F/\partial d$, with the Einstein relation between the mobility $b$ and diffusion $B$.

$B$ can be estimated as $\nu a^2\exp(-F_B/kT)$ where $\nu$ is the characteristic atomic frequency ($\sim 10^{13}$ s$^{-1}$), $a$ is the characteristic interatomic distance, $F_B$ is the kinetic phase transformation barrier, $k$ is Boltzmann's constant, and $T$ is the temperature. Assuming $B$ of the order of the diffusion coefficient $D$ of species dominating whisker growth, it can be estimated based on the available data; otherwise $B$ remains a parameter of this theory.

The averaged description in Eqs. (\ref{eq:mobility}) can apply as long as it is not affected by rare events terminating or exponentially slowing whisker growth. These events take place when whisker tips reach random local regions of abnormally low fields. They present barriers to whisker growth, since the electrostatic energy gain there is suppressed, while surface related energy loss remains. The statistics of these rear events and its corresponding distribution of whisker lengths are described in Sec. \ref{sec:stat} below.

\subsection{Limiting cases}\label{sec:limit}

Integrating Eq. (\ref{eq:mobility}) with $F=-F_E$ (i. e. neglecting surface energy far enough from the nucleation barrier) and $F_E$ from Eqs. (\ref{eq:Egain}) and (\ref{eq:WE1}), and treating $\Lambda$ as a constant, yields
\begin{eqnarray}\label{eq:t0}
h\approx &&\frac{h_0}{1-t/t_0},\quad d\approx\sqrt{d_0^2+(h^2-h_0^2)/\Lambda },\\
&& t_0\equiv \frac{3\Lambda}{b\varepsilon E_0^2h_0}\quad {\rm when}\quad h\ll L,\nonumber
\end{eqnarray}
and
\begin{eqnarray}\label{eq:tL}
h=L\frac{t}{t_L}, \quad d\approx L/\sqrt{\Lambda},\\ t_L\equiv \frac{3\Lambda}{b\varepsilon E_0^2L}\quad {\rm when}\quad r_c\gg h\gg L.\nonumber
\end{eqnarray}
We recall that underlying these results is the
approximation $\Lambda \gg 1$. Finally, in  the region of yet larger lengths, $h\gg r_c$, the whiskers will grow uniformly, as predicted by Eq. (\ref{eq:tL}) where $E_0^2$ is replaced with $\langle E_T^2\rangle$.

The result in Eq. (\ref{eq:tL}) changes numerically when more accurate expressions (\ref{eq:finalFE}) and (\ref{eq:avfinal}) from Appendices are used for the free energy $F_E$. The growth rate from Eq. (\ref{eq:tL}) acquires then an additional coefficient to become
\begin{equation}\label{eq:tLnew}
 \frac{dh}{dt}\approx \frac{L}{t_L}\ln\left(\frac{4t}{t_L}\right)\quad {\rm with}\quad t_L\equiv \frac{32\pi (\Lambda +4/3)}{b\varepsilon E_0^2L}.
 \end{equation}

\begin{figure}[bth]
\includegraphics[width=0.45\textwidth]{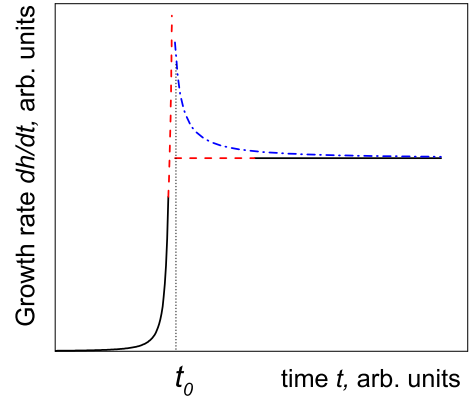}
\caption{Sketch of the whisker growth rates vs. time. Solid lines represent the two limiting cases given by Eqs. (\ref{eq:t0}) and (\ref{eq:tL}) within the domains of their applicability ($h\ll L$ and $h\gg L$ respectively). The dashed curves are formal solutions of Eqs. (\ref{eq:t0}) and (\ref{eq:tL}) beyond that domains where they must be sewed. The dash-dotted line shows a hypothetical sewing that would fit the observations of Ref. \onlinecite{jadhav2010}. \label{Fig:grate}}
\end{figure}

Fig. \ref{Fig:grate} presents the predicted temporal dependence of whisker growth rate where the limiting results must be sewed at $h\sim L$, i. e. $t\sim t_0$. In that poorly described region, the two predicted limiting case rates match in the order of magnitude. Indeed, setting $h\sim L$ in Eq. (\ref{eq:t0}) yields $1-t/t_0\sim h/L$, which results in the same $dh/dt$ as provided by Eq. (\ref{eq:tL}).

However, the details of the latter sewing remain unknown. Assuming a hump in the sewing region would make Fig. \ref{Fig:grate} resembling the real time data from Ref. \onlinecite{jadhav2010}. Even without that hump, Fig. \ref{Fig:grate} reproduces the observation of many authors that whisker growth starts abruptly from some time instance ($t_0$ in this work notations) to continue at constant rate; the numerical estimates are given in Sec. \ref{sec:num}.

The above description of whisker growth is essentially one-dimensional. A phenomenon of whisker growth in a labyrinth of  random electric fields is beyond the scope of this work.

\section{Long range diffusion}\label{sec:longr}
The fact that whiskers grow from their roots without forming any surrounding dents is commonly interpreted as a result of long range uniform drift of material towards whisker roots. That interpretation was experimentally verified (see e. g. Refs. \onlinecite{crandall, woodrow2006}).

The drift necessity follows naturally from the electrostatic theory here. Indeed, taking into account that $E\propto n$, the electrostatic energy related to surface charge density $n$ is proportional to the surface integral $\int n^2dA$. The latter must be a minimum under the condition of charge conservation, $\int ndA=const$. It is straightforward to see that the required conditional minimum takes place when $n=const$, i. e. surface charge is distributed uniformly within its occupied domain. Assuming that charges are tightly pinned to the surface material, the system should maintain uniform material density; hence, long range drift.

The integral laws of minimum energy and charge conservation above do not specify the underlying forces.  In addition to the long range Coulomb, there must be some short range forces tightly binding surface charges to surface material and making it move along. A hydrodynamic drag appears to be a conceivable mechanism of such coupling. It is qualitatively similar to a flow caused by a dense enough array of particles pushed through a viscous fluid. This hypothesis remains to be verified at a more quantitative level.

\section{Whisker statistics}\label{sec:stat}
As explained in the end of Sec. \ref{sec:genfor} above, whisker growth is blocked in the local regions of anomaly low electric field. Because these regions have random locations, this will result in broad statistical distributions of whisker lengths. The observed distributions of this kind were best approximated as log-normal. \cite{fang2006,panashchenko2009,susan2013} Their analytical form is derived here.

\begin{figure}[bth]
\includegraphics[width=0.35\textwidth]{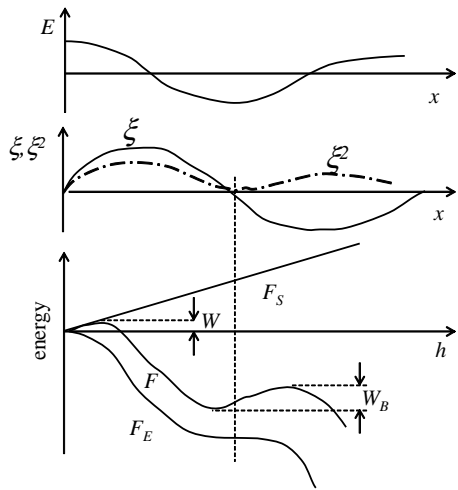}
\caption{Coordinate dependencies of random electric field $E(h)$ (top), random functional $\xi\equiv\int _0^hE(x)dx$ and its square (middle), and free energies: surface contribution $F_S$, electrostatic contribution $F_E$, and their sum $F=F_E+F_S$. The vertical dash line marks the point of $\xi =0$ of $F_E$ flattening that gives rise to a barrier $W_B$ blocking the whisker growth. The barrier $W$ is the same as in Fig. \ref{Fig:F}.\label{Fig:stopper}}
\end{figure}

As shown in the Appendix below [Eq. (\ref{eq:finalFE})], the electrostatic energy gain of a whisker in a nonuniform field $E(x)$ can be presented  in the form
\begin{equation}\label{eq:Frand}
F_E(h)=-\frac{1}{4\Lambda}\int _0^h\xi ^2dx, \quad \xi (x)\equiv \int _0^xE(x')dx'.
\end{equation}
Because fields $E$ are random, $F_E(h)$ becomes a random functional giving rise to barriers in the total free energy $F=F_E+F_S$ as illustrated in Fig. \ref{Fig:stopper}.

Taking into account surface energy, $F_S=\sigma\pi dh$, the barrier condition $dF/dh=0$ reduces to $\xi ^2(h)=4\Lambda\sigma\pi d$ much below its average, $\xi ^2\ll \langle\xi ^2\rangle$. Such $\xi ^2$ correspond to flat portions in the dependence $F_E(h)$ where the surface contribution slope $\sigma\pi d$ dominates free energy; this is reflected in Fig. \ref{Fig:stopper}.

The typical barriers block whisker growth during the time of experiment. Indeed, they correspond to the characteristic length scale of field variation of the order of $h$ [see the discussion after Eq. (\ref{eq:WE1})], which leads to the  estimate $W_B\sim \pi dh\sigma\gtrsim 100$ eV. On the other hand, we assume long enough times of experiment, during which the average growth rates from Eqs. (\ref{eq:t0}), (\ref{eq:tL}) allow whisker lengths to reach the regions of blocking barriers.  

With the latter assumption, the distribution of whisker lengths can be approximated by the distribution of coordinates of their blocking barriers. It is given by the probability density $g(\xi ^2)$ for the random quantity $\xi ^2$ in a close proximity of  $\xi ^2 =0$. As a square of integral over large distances, $\xi ^2$ can be thought of as a sum of large number of random contributions. According to the central limit theorem, such random quantities are described by the Gaussian distribution,
\begin{equation}\label{eq:gxi}g(\xi ^2)=\frac{1}{\sqrt{2\pi \Delta}} \exp\left[-\frac{(\xi ^2-\langle\xi ^2\rangle )^2}{2\Delta}\right],\end{equation}
where angular brackets represent averaging and $\Delta =\langle \xi ^4\rangle -(\langle\xi ^2\rangle )^2$ is the dispersion.

\begin{figure}[t]
\includegraphics[width=0.52\textwidth]{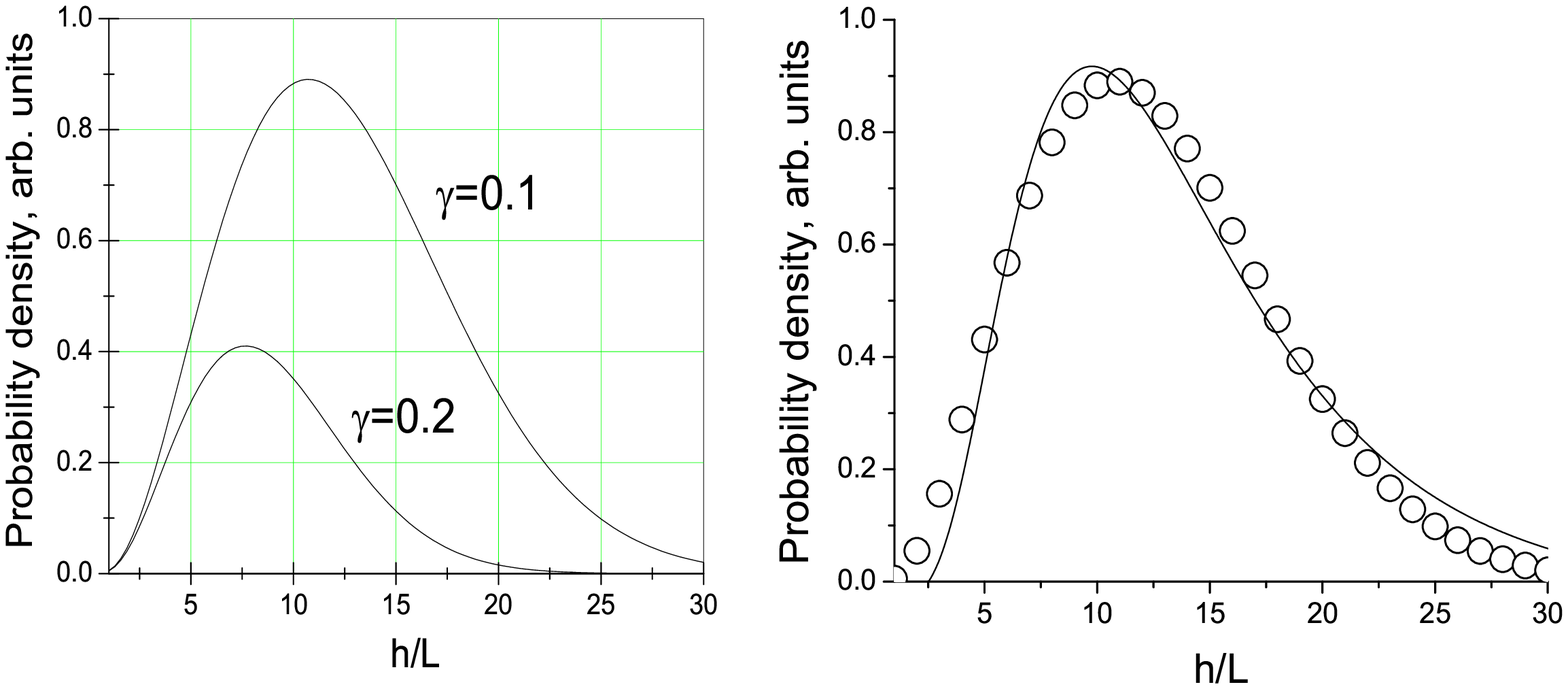}
\caption{Left: Probabilistic distributions of Eq. (\ref{eq:dist}) for two different values of the numerical parameter $\gamma$. Right: Comparison between probabilistic distribution of Eq. (\ref{eq:dist}) (open circles) and a log-normal fitting curve (solid line). \label{Fig:stat}}
\end{figure}

The momenta $\langle\xi ^2\rangle$ and $\langle \xi ^4\rangle$ depend on distance $h$ to the metal surface. They can be evaluated based on statistical properties of surface charge fluctuations, which we assume to be uncorrelated at distances exceeding the patch size $L$. 
Using the corresponding results for $\langle\xi ^2\rangle$ and $\langle \xi ^4\rangle$ from Appendix \ref{sec:xi}, the probabilistic distribution of whisker lengths, takes the form
\begin{eqnarray}\label{eq:dist}&&g(h) = \frac{1}{\sqrt{2\pi \Delta (h)}}\exp \left[-\frac{(\langle\xi ^2(h)\rangle  )^2}{2\Delta (h)}\right] \\
&=& \beta\frac{h}{L} \exp\left\{-\gamma\left[\frac{h}{L}\ln\left(\frac{[1+\sqrt{1+(h/L)^2}]^2}{4\sqrt{1+(h/L)^2}}\right)\right]^2\right\}.\nonumber\end{eqnarray}
Here the unknown numerical coefficients $\beta$ and $\gamma$ should not be too different from unity.

The distribution of Eq. (\ref{eq:dist}) is illustrated in Fig. \ref{Fig:stat}. It is a maximum at $h/L\approx\exp (1/\sqrt{2\gamma})$ with half-width $\delta(h/L)\approx (\ln 2)(2\gamma )^{-1/4}\exp[1/(4\gamma )]$. The log-normal distribution fit is demonstrated in the same figure is in agreement with the experimental statistics. \cite{fang2006,panashchenko2009,susan2013}

Two significant limitations underlying Eq. (\ref{eq:dist}) are that of the 1D model of whisker evolution and the infinitely blocking barriers. Relaxing either of them would allow additional whisker growth.

\section{Numerical estimates and discussion}\label{sec:num}
Numerical estimates of nucleation parameters were given in Sec. \ref{sec:fin}). Here we estimate the parameters of whisker growth. Using the diffusion coefficient for tin \cite{woodrow2006} $D\sim 10^{-18}$ cm$^2$s$^{-1}$ and assuming $D=B$ gives for the mobility $b\sim 4\times 10^{-5}$ cm$^2$s$^{-1}$erg$^{-1}$. Substituting the latter along with $E_0\sim 1$ MV/cm, $h_0\sim 10$ nm, and $\Lambda =1-3$, Eq. (\ref{eq:t0}) yields $t_0\sim (0.3-1)\times 10^5$ s. This is in fair agreement with the data \cite{jadhav2010,susan2013} on the incubation period preceding constant growth rate (cf. Fig. \ref{Fig:grate}).

For a reasonable patch size $L\sim 3$ $\mu$m, the growth rate predicted by Eq. (\ref{eq:tL}) turns out to be by two orders of magnitude higher than the observed $dh/dt\sim 1$ {\AA}/s. This discrepancy can be due to oversimplifications related to the 1D random field treatment.
Using more accurate Eq. (\ref{eq:tLnew}) decreases $dh/dt$ by at least one order of magnitude bringing it closer to the data.

The latest stage of whiskers growth above critical length $r_c$ is described by the characteristic time that is by the factor $E_0^2/\langle E_T^2\rangle\sim 10^5$ longer than $\tau _L$, i. e. $t_T\sim 10^7-10^8$ s. This agrees with the observations that whiskers typically stop growing in a year or so. (Blocking barriers introduced in Sec. \ref{sec:stat} can provide an alternative explanation of that observation.)

Reasonable agreement with statistical data can be demonstrated by using again $L\sim 3$ $\mu$m and assuming $\gamma\sim 0.1-0.2$. In that case, the curves in Fig. \ref{Fig:stat} become close to the published statistical data. \cite{fang2006,panashchenko2009,susan2013}

Because a significant effort was spent to understand metal whiskers in terms of mechanical stresses, recrystallization, dislocations, etc., it should be noted that the present theory does not rule out these factors. Furthermore, they can be a part of the picture presented in two major aspects. First, they can create local spots of unfavorable energy configurations capable of relaxing through the mechanism of field induced nucleation. That belongs in the general settings of inhomogeneous nucleation strongly facilitated by imperfections and inhomogenieties.  Secondly, local spots of stress, grain boundaries, or dislocations can be sources of surface charge leading to the field induced nucleation as discussed in Sec. \ref{sec:field} above. 

The above estimates lead to the following scenario of whisker evolution. (i) Stage 1: whiskers nucleate in a sub-second to days time interval (reflecting fluctuations in nucleation barriers due to the local field fluctuations); their dimensions upon nucleation are $h\sim 10$ nm and $d\sim 1$ nm, with predominant orientation perpendicular to the metal surface. (ii) Stage 2: Whiskers grow up to the patch size, say $L\sim 3-10$ $\mu$m, more or less perpendicular to the surface, with some deviations especially towards patch edges. This takes a much longer time $t_0\sim 10^4-10^5$ s that can be experimentally identified as the whisker incubation time. The growth rate at this stage is very low for almost entire time interval $t_0$, with drastic acceleration in the nearest proximity of $t_0$ (see Fig. \ref{Fig:grate}). (iii) Stage 3: Whiskers grow way above patch size $L\sim 3-10$ $\mu$m at constant rate $L/t_L$ possibly with some degree of winding or kinking (beyond the current theory). At this stage, random field configurations induced by uncorrelated patch charges make growth rates of individual whiskers fluctuating, some of them blocked. The random distribution of blocking barriers determines the statistical distribution of whisker lengths. (iv) Stage 4: If whiskers grow above lengths where feeding by thermal radiation dominates, they evolve further in lateral directions parallel to the metal surface.

\section{Conclusions}\label{sec:concl}
The above theory describes metal whickers as a result of metal nucleation and growth in random electric fields induced by charged patches on metal surfaces. The underlying approaches are typical of the physics of phase transitions and disordered systems. This work presents the first whisker theory yielding simple analytical results more or less consistent with the observations. The successes, the remaining questions, and possible experimental verifications of this theory are summarized next.

\subsection{What is understood}\label{sec:whatund}

1) Why whiskers are metallic: high (metallic) electric polarizability is required for sufficient energy gain due to whisker formation in external (surface) electric fields.\\
2) Why whiskers grow more or less perpendicular to the surface: such are the dominating directions of the surface electric field.\\
3) Why whisker parameters are broadly statistically distributed: this reflects fluctuations in metal surface fields induced by mutually uncorrelated charged patches.\\
4) Correlation between whiskers and versatile morphology factors, such as (i) grains whose orientation is different from the major
orientation of the tin film, (ii) dislocations and dislocation loops, and (iii) mechanical stresses capable of surface buckling, surface contaminations; all related to local surface charges and their induced electric fields. Some metals are more prone to develop whiskers because they can easier form charged patches by absorbing ions, and creating dislocations, grain boundaries, or stresses.\\
5) Why external electric bias can significantly accelerate whisker growth: external electric fields increase the nucleation and growth rates.\\
6) Why the characteristic whisker evolution follows a certain pattern: long incubation period followed by almost constant growth rate that eventually saturates. The predicted incubation time and subsequent growth rate agree with the observations. \\
7) Why whisker parameters are broadly distributed statistically. The predicted distribution of whisker lengths is close to the observed log-normal statistics.
\subsection{What is not understood}\label{sec:whatnot}
1) The microscopic nature of whiskers, their correlation with specific surface defects, chemical aspects of whisker development.\\
2) The role of whisker crystalline structure in their evolution process.\\
3) Whisker growth in 3D random electric field. This includes whisker winding and kinking.\\
4) Possible role of surface (or grain boundary) diffusion limiting whisker growth.\\
5) Possible hydrodynamic drag moving surface material uniformly  along with ions.\\
6) Inter-whisker interactions limiting their concentration and affecting growth.\\

\subsection{Possible experimental verification}

The predicted dependencies of nucleation and growth kinetics vs. electric field, temperature, and controlled contamination could be verified experimentally.

1) Whisker nucleation and growth in external electric fields. This can be attempted in either flat capacitor configuration of for a whisker inside SEM where the electric field is readily controlled. In both cases, care should be taken to avoid significant Joule heating and/or electron drag effects, i. e. using voltage rather than current power source.\\
2) Whisker nucleation and growth under controlled contamination of metal surface with solutions of charged nano particles. \\
3) Whisker nucleation and growth under the conditions of strong surface electric fields induced by surface plasmon polariton excitations. \cite{maier2007} This technique could be used for controlled growth of metal nanowires of desirable parameters on  metal surfaces.\\

In the end, it should be noted that this work presents rather a sketch of theory in its infancy, pointing at important factors and providing estimates, yet not enough developed to quantitatively describe whisker evolution and statistics in a random electric field. Further effort is called upon to develop this approach.
\section*{Acknowledgement}
Discussions with D. Shvydka, A. V. Subashiev, I. V. Karpov, E. Chason, and D. Susan are greatly appreciated.
\appendix
\section{Nonuniform polarization}\label{sec:append}
Following the approach in Ref. \onlinecite{landau1984}, p. 17, the electrostatic energy gain of a rectilinear metal filament of length $h$ and radius $a\ll h$ in a {\it nonuniform} electric field $E_(x)$ is given by
\begin{equation}\label{eq:FE}F_E=(1/2)\int _0^h \varphi (x)\tau (x)dx\end{equation}
where $\tau$ is the field induced charge density and $\varphi (x)=-\int _0^xEdx$ is the electric potential. The condition of constant potential on the surface of the filament is
\begin{eqnarray}\label{eq:pot}
&&-\int Edx +\frac{1}{2\pi}\int_0^{2\pi} \int_0^h\frac{\tau (x')dx'd\phi}{R}=0,\\
&&R=\sqrt{(x-x')^2+4a^2\sin ^2(\phi /2)}\nonumber
\end{eqnarray}
where $\phi$ is the angle between planes passing through the axis of the cylinder and through two points on its surface
at a distance $R$ apart. We divide the integral into two parts, putting $\tau (x') = \tau (x) + [\tau (x') -\tau (x)]$. Since $h\gg a$, we
have for points not too near the ends of the rod,
\begin{equation}\frac{\tau (x)}{2\pi}\int\int\frac{dx'd\phi}{R}\approx \tau (x)\log\frac{4(l^2-x^2)}{a^2}.
\end{equation}
Thus,
\begin{equation}\label{eq:tau}
\int _0^xEdx=\tau (x)\log\frac{4(hx-x^2)}{a^2}+\int _0^h\frac{\tau (x')-\tau (x)}{|x'-x|}dx'.
\end{equation}
It follows then that $\tau (x)$ is almost proportional (with logarithmic accuracy) to $\int _0^xEdx$; it can be sought in the form $\tau =A\int _0^xEdx$ that should be substituted into Eq. (\ref{eq:tau}). Noting that the integral in Eq. (\ref{eq:tau}) is dominated by the proximity $x'=x$ and representing the integrand numerator as $E(x)(x-x')$, one finally gets
\begin{equation}\label{eq:fintau}\tau (x)=\frac{\int _0^xE(x)dx}{\log[4(hx-x^2)/a^2]-2}.\end{equation}

Substituting the latter and assuming as in the body of the text that $\Lambda\gg 1$, the electrostatic free energy becomes
\begin{equation}\label{eq:finalFE}
F_E(h)=-\frac{1}{4(\Lambda +4/3)}\int_0^h\left(\int _0^xE(x')dx'\right)^2dx
\end{equation}
where $\Lambda $ is defined in Eq. (\ref{eq:ncyl}).
This result reproduces the above used $F_E\propto -h^3$ and $F_E\propto -h$ for the cases of $h\ll L,\quad E=cosnt$ and $h\gg L, \quad E\propto 1/x$ respectively; also, it justifies the claim in Sec. \ref{sec:highch} that the point charge caused polarization does not impact the electrostatic energy gain.
\section{Random variable $\xi ^2$}\label{sec:xi}
Here we evaluate the parameters of statistical distribution of Eq. (\ref{eq:gxi}), $\langle\xi ^2\rangle$ and $\Delta=\langle \xi ^4\rangle -(\langle\xi ^2\rangle )^2$. The required averaging gives
\begin{eqnarray}\label{eq:dispdef}
\langle\xi ^2\rangle &=&\left\langle \left(\int _0^hdr\int d^2\rho\frac{en(\boldsymbol\rho )r}{(r^2+\rho ^2)^{3/2}}\right)^2\right\rangle \\
&=&\int d^2\rho\int d^2\rho '\frac{e^2\langle n(\boldsymbol\rho)n (\boldsymbol\rho ')\rangle}{(\rho ^2+h^2)^{1/2}(\rho '^2+h^2)^{1/2}}.\nonumber
\end{eqnarray}
Here $\boldsymbol\rho$ and $\boldsymbol\rho '$ are two-dimensional radius-vectors in the plane of surface charge; $d^2\rho$ and $d^2\rho '$ are elemental areas. Assuming uncorrelated random charge patches, one can write
\begin{equation}\label{eq:correlator}
e^2\langle n(\boldsymbol\rho)n (\boldsymbol\rho ')\rangle=C_1\delta (\boldsymbol\rho -\boldsymbol\rho ')
\end{equation}
where $\delta$ stands for the Dirac delta-function, and $C_1$ is a constant. Its order of magnitude estimate is $C_1\sim \langle (neL)^2\rangle$. The delta-function representation remains adequate when $\rho\gg L$ where $L$ is the linear dimension of a charged patch. Substituting the latter correlation function into Eq. (\ref{eq:dispdef}) and performing integration over $\rho$ from $L$ to $\infty$ yields
\begin{equation}\label{eq:avfinal}
\langle\xi ^2\rangle =2\pi C_1\ln\left[\frac{(L+\sqrt{L^2+h^2})^2}{4L\sqrt{L^2+h^2}}\right].
\end{equation}
For the region of $h\gg L$, Eq. (\ref{eq:avfinal}) simplifies to $\langle\xi ^2\rangle =C_1\ln (h/4L)$.

Similarly, $\langle\xi ^4\rangle$ can be reduced to the integral over area elements $d^2\rho d^2\rho ' d^2\rho ''d^2\rho ''' $ with the integrand
\begin{equation}\nonumber
\frac{\langle n(\boldsymbol\rho )n (\boldsymbol\rho ')n (\boldsymbol\rho '')n (\boldsymbol\rho ''')\rangle }{(\rho ^2+h^2)^{1/2}(\rho '^2+h^2)^{1/2}(\rho ''^2+h^2)^{1/2}(\rho '''^2+h^2)^{1/2}}
\end{equation}
Here, $\langle n\rangle =0$; hence, finite contributions to the integral arise from the product of two pair averages $e^2\langle n(\boldsymbol\rho)n (\boldsymbol\rho ')\rangle$ given in Eq. (\ref{eq:correlator}), and from the correlation function $e^4\langle n(\boldsymbol\rho)n (\boldsymbol\rho ')n (\boldsymbol\rho '')n (\boldsymbol\rho ''')\rangle=C_2\delta (\boldsymbol\rho -\boldsymbol\rho ')\delta (\boldsymbol\rho -\boldsymbol\rho '')\delta (\boldsymbol\rho -\boldsymbol\rho ''')$. The former product cancels out with $\langle\xi ^2\rangle ^2$ in the definition of dispersion $\Delta$. The latter term yields
\begin{equation}\label{eq:disfin}
\Delta =2\pi C_2h^{-2}.
\end{equation}
The order of magnitude estimate for the coefficient is $C_2\sim |ne|^4L^6$. Therefore, the dispersion is relatively small,
\begin{equation}\label{eq:relative}
\frac{(\langle\xi ^2\rangle )^2}{\Delta}\sim \left[\frac{h}{L}\ln\left(\frac{h}{4L}\right)\right]^2\gg 1.
\end{equation}

\end{document}